\documentstyle[twoside,fleqn,espcrc2]{article}

\newcommand{\ttbs}{\char'134}
\newcommand{\AmS}{{\protect\the\textfont2
  A\kern-.1667em\lower.5ex\hbox{M}\kern-.125emS}}

\title{%
\begin{flushright}
{\small WU-HEP-92-10, IPS Research Report No.92-28, HUPD-9218}\\
\end{flushright}
Autocorrelation in Updating Pure SU(3) Lattice Gauge Theory
by the use of Overrelaxed Algorithms
}
\author{\vskip5mm
\begin{center}
        {\large QCD\_TARO Collaboration }\\
\end{center}
        \vskip5mm
        K.Akemi
           \address{Computational Science Research Laboratory,
                    Fujitsu Limited, Ota-ku, Tokyo 144, Japan},
        Ph.deForcrand
           \address{IPS,ETH-Z\"urich, CH-8092 Z\"urich, Switzerland},
        M.Fujisaki$^{\rm a}$,
        T.Hashimoto
           \address{Dept. of Applied Physics, Fukui University,
                    Fukui 910, Japan},
        H.C.Hege
           \address{ZIB, D-1000 Berlin 31, Germany},
        S.Hioki
           \address{Dept. of Physics, Hiroshima University,
                    Higashi-Hiroshima 724, Japan},
        O.Miyamura$^{\rm e}$,
        A.Nakamura
           \address{Dept. of Physics, Waseda University,
                    Tokyo 169, Japan},
        M.Okuda$^{\rm a}$,
        I.O.Stamatescu
           \address{FESt Heidelberg and
                 Institut f\"ur Theoretische Physik, Universit\"at Heidelberg,
                    D-6900 Heidelberg, Germany},
        Y.Tago$^{\rm a}$
        and
        T.Takaishi$^{\rm e}$
       }

\begin{document}

\begin{abstract}
  We measure the sweep-to-sweep autocorrelations of blocked loops
below and above the deconfinement transition for SU(3) on a $16^4$ lattice
using 20000-140000 Monte-Carlo updating sweeps.
 A divergence of the autocorrelation time toward the critical $\beta$ is seen
at high blocking levels.
 The peak is near $\beta$ = 6.33 where we observe  440 $\pm$ 210
for the autocorrelation time
of $1\times 1$ Wilson loop on $2^4$ blocked lattice.
  The mixing of 7 Brown-Woch overrelaxation steps
followed by one pseudo-heat-bath step
appears optimal to reduce the autocorrelation time below the critical $\beta$.
  Above the critical $\beta$, however, no clear difference between these two
algorithms can be seen and the system decorrelates rather fast.
\end{abstract}

\maketitle

\section{Introduction}

 In the Monte Carlo evaluation of physical observables,
we should gather independent samples to get
statistically meaningful results.
 We generate a sequence of configurations and
then we choose samples separated by some sweeps to ensure their independence.
 This separation should be larger than the Monte Carlo autocorrelation
time $\tau$.
Otherwise statistical errors should be corrected to take correlations
into account.  It is crucial in lattice QCD to estimate correctly these
errors and the underlying autocorrelation time.
 In this paper we report autocorrelation measurements in
pure SU(3) lattice gauge theory.\cite{AC}

 The aims of our study are
\begin{enumerate}
 \item to extract autocorrelation time
 \item to see operator dependence of autocorrelation,i.e. blocked loops
 \item to see $\beta$ dependence of autocorrelation
       which tells us some information
       about the critical slowing down toward the continuum limit.
\end{enumerate}

\section{Overrelaxation, Blocking scheme and Numerical simulation}

 To update the gauge field here we use Cabbibo-Marinari\cite{CM}
 pseudo-heat-bath(PHB)
with 3 SU(2) subgroups and to perform the overrelaxation(OR) we adopt
Brown-Woch\cite{BW} microcanonical method
with 2 randomly chosen SU(2) subgroups.

We introduce a parameter $K$ which denotes the mixing ratio
between OR and PHB, namely we choose OR or PHB stochastically
with the probabilities $K/(K+1)$ and $1/(K+1)$.

 For blocking we use the Swendsen\cite{SW} factor 2 blocking scheme.
For details and results of our MCRG study,
see Ref.\cite{RG}

 We work on a $16^4$ lattice and
 20000-140000 sweeps are performed at all $\beta$.
All data are $K=9$ case except for $\beta$=6.3 and 6.8.
At $\beta$=6.3 and 6.8, $K$=0, 3, 5, 7, 9, 12 were studied
to check the efficiency of OR algorithm.

 After a few thousands of sweeps for thermalization, we measure operators:
W11, W12, W22, Wchair, Wsofa and Wtwist at every blocking level from 1 to 4
(i.e. from $16^4$ to $2^4$).
Blocking is performed every 10 sweeps.
 All our simulations have been performed on 64-node AP1000 at Fujitsu.

\section{Results}

 In Fig.1,
we show typical Monte Carlo history of W11(1$\times$1 wilson loop)
at blocking levels 2-4 at $\beta$=6.38.

\begin{figure}[t]
\unitlength 1.0mm

\end{figure}

 In this case no clear signal of sweep-to-sweep correlation at low
blocking level can be seen whereas at high blocking level
some structure emerges which we cannot see at lower levels.

 Then we define the autocorrelation function $\rho(t)$ of observable $O$ as
\begin{equation}
\rho(t) = { {\langle(O(i)-O_A)(O(i+t)-O_B)\rangle} \over
      {\sqrt{\langle(O(i)-O_A)^2\rangle\langle(O(i+t)-O_B)^2\rangle}} }
\end{equation}
where $\langle\ \rangle$ denotes average over $i$ and
$O_A = \langle O(i)\rangle$ and $O_B = \langle O(i+t)\rangle$.

 Fig.2 shows the autocorrelation function $\rho(t)$ for t=0-1000 sweeps
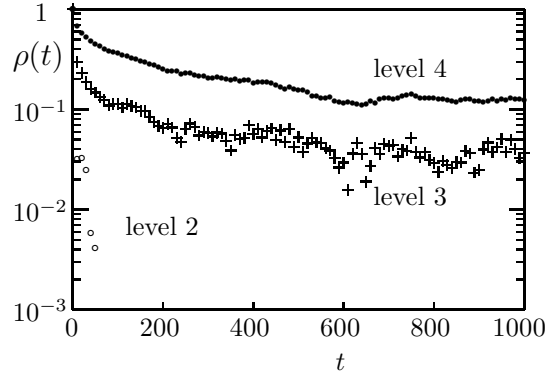
for the same data as in Fig.1.
 In Fig.2 we see long range autocorrelation, as stated above, at high
blocking level.

\begin{figure}[t]
\unitlength 1.0mm

\end{figure}

 This tendency is very natural.
 Since W11 at high blocking level has effectively large area, then
more Monte Carlo sweeps are needed to alter its value.
 In other words, W11 at high blocking level
is not easily influenced by short range fluctuations whereas
at lower level it is.

 Now we introduce integrated autocorrelation time $\tau_{int}$
defined as:
\begin{equation}
 \tau_{int} = \rho(0) + 2 \sum_{t=1}^{N} \rho(t) {{N-t} \over N}
\end{equation}

 In this analysis we choose $N$=1000, so that
 $\tau_{int}$ can extract long range information.
 For the error estimation of correlated samples,
$\tau_{int}$ is the relevant quantity and enters in our
final error analysis.

 We have done this analysis not only for W11 but also for other measured
loops such as W12, W22, Wsofa etc.
 There exist, of course, differences among them about
autocorrelation function and time, however, we found that
for a given blocking level, these differences are not significant.
 So from now on, we concentrate on W11 at level=4.

 As stated above, we have checked the efficiency of OR algorithm changing
the mixing parameter $K$.

 In Fig.3(a), $\rho(t)$ for $t$=0-600
at $\beta$=6.3 for $K$=0,3,7 and 9 are displayed.
 In this figure, $K$=7 seems to work well to reduce the autocorrelation.
To see this quantitatively, we show $\tau_{int}$ defined by eq.(2) for
several values of $K$ at $\beta$=6.3 in Fig.3(b).
We can see the efficiency of $K$=7 case also in this figure.

{}From this analysis, we can extract the typical autocorrelation time
of W11 at level=4 at $\beta$=6.3 such as:
about 300 sweeps for PHB, about 100 sweeps for OR at $K$=7.
  In ref.\cite{DF} Decker and de Forcrand give similar numbers.
 These numbers are somewhat larger than those obtained by other authors.
 This is because we are using blocked loops as observables.
To see the long range autocorrelation, extended operators look suitable
because small operators are easily influenced by fluctuations.

\begin{figure}[t]
\unitlength 1.0mm

\end{figure}

When we want to include error bars, we must be careful to control these
long range autocorrelation effects.

Let $\delta\rho_0(t)$ be the error on $\rho(t)$
calculated as if samples were independent.
Then if we consider the dependence between samples, the error is modified as
\begin{equation}
 (\delta\rho(t))^2 = (\delta\rho_0(t))^2 \tau_{int}.
\end{equation}

\begin{figure}[t]
\unitlength 1.0mm
%
\end{figure}

This error on $\rho(t)$ generates an error on the extracted autocorrelation
times.
 We estimate it by
\begin{equation}
 \delta\tau_{int} = {\delta\rho(\tau_{int}) \over \rho(\tau_{int})} \tau_{int}.
\end{equation}

To see the definition dependence of error analysis, we also evaluate
errors using jackknife method.
To perform jackknife analysis for error estimates of autocorrelation function,
 we first divide the sample into bins.
We get $\delta\rho(t)$ and then we can estimate the error
of autocorrelation time following eq.(4).
It is thought that if we double the size of a bin,
error increases roughly by a factor $\sqrt{2}$ when size of a bin is smaller
than the autocorrelation time and then levels off when size of a bin exceeds
the autocorrelation time.
 In our analysis there are many cases when the error won't level off.
In such cases the error on the error of autocorrelation function is very large.

We show the results of this jackknife study also in Fig.3(b).
The errorbar with full circle comes from using eq.(3) and that with open circle
comes from jackknife analysis.
In all cases of $K$, jackknife method estimates smaller errorbars than
eq.(3).

If we include the error and consider again the efficiency of
OR algorithm, we can only say that $K$=7 looks optimal.
To adopt OR algorithm, it is important to choose the optimal value
of $K$ to update the system efficiently. The optimum range of $K$
is surprisingly narrow.

We have performed the same analysis also at $\beta$=6.8.
At this $\beta$, $\tau_{int}$ is rather small for any $K$
(typically less than 50 sweeps)
 and no clear advantage of OR can be seen.


\begin{figure}[t]
\unitlength 1.0mm
\begin{picture}(60,60)(-10,-10)
\def\xw{60.000000} \def\yw{40.000000}
\put(14.1,6.2){$\uparrow$}
\put(11,3){6.10}
\put(25,5){$\leftarrow$}
\put(29,4.6){6.80}
\put(31,12){$\uparrow$}
\put(28.3,8){6.43}
\put(50,31){6.33}
\put(27,-8){$t$}
\put(-8,32.5){\large $\rho(t)$}
\put(-10,-16){Fig.4(a)\ \ $\rho(t)$ vs. $t$, for several $\beta$ at $K$=9.}
\put(0.000000,40.000000){\circle*{0.700000}}
\put(1.000000,35.416508){\circle*{0.700000}}
\put(2.000000,33.908920){\circle*{0.700000}}
\put(3.000000,32.568733){\circle*{0.700000}}
\put(4.000000,31.529573){\circle*{0.700000}}
\put(5.000000,31.020294){\circle*{0.700000}}
\put(6.000000,29.494841){\circle*{0.700000}}
\put(7.000000,28.186934){\circle*{0.700000}}
\put(8.000000,26.580133){\circle*{0.700000}}
\put(9.000000,23.685066){\circle*{0.700000}}
\put(10.000000,26.144133){\circle*{0.700000}}
\put(11.000000,25.169867){\circle*{0.700000}}
\put(12.000000,20.926666){\circle*{0.700000}}
\put(13.000000,20.334133){\circle*{0.700000}}
\put(14.000000,23.256134){\circle*{0.700000}}
\put(15.000000,22.746799){\circle*{0.700000}}
\put(16.000000,23.076000){\circle*{0.700000}}
\put(17.000000,23.356400){\circle*{0.700000}}
\put(18.000000,22.470266){\circle*{0.700000}}
\put(19.000000,23.944134){\circle*{0.700000}}
\put(20.000000,20.414667){\circle*{0.700000}}
\put(21.000000,14.997466){\circle*{0.700000}}
\put(22.000000,16.463467){\circle*{0.700000}}
\put(23.000000,16.906933){\circle*{0.700000}}
\put(24.000000,5.702534){\circle*{0.700000}}
\put(25.000000,17.011600){\circle*{0.700000}}
\put(0.000000,40.000000){\circle*{0.700000}}
\put(1.000000,37.857014){\circle*{0.700000}}
\put(2.000000,37.020439){\circle*{0.700000}}
\put(3.000000,36.384228){\circle*{0.700000}}
\put(4.000000,35.807522){\circle*{0.700000}}
\put(5.000000,35.453560){\circle*{0.700000}}
\put(6.000000,34.990692){\circle*{0.700000}}
\put(7.000000,34.604279){\circle*{0.700000}}
\put(8.000000,34.203678){\circle*{0.700000}}
\put(9.000000,33.897560){\circle*{0.700000}}
\put(10.000000,33.828785){\circle*{0.700000}}
\put(11.000000,33.542213){\circle*{0.700000}}
\put(12.000000,33.265614){\circle*{0.700000}}
\put(13.000000,33.376972){\circle*{0.700000}}
\put(14.000000,33.229839){\circle*{0.700000}}
\put(15.000000,32.811493){\circle*{0.700000}}
\put(16.000000,32.700401){\circle*{0.700000}}
\put(17.000000,32.625572){\circle*{0.700000}}
\put(18.000000,32.565266){\circle*{0.700000}}
\put(19.000000,32.571789){\circle*{0.700000}}
\put(20.000000,32.314293){\circle*{0.700000}}
\put(21.000000,31.946814){\circle*{0.700000}}
\put(22.000000,31.718613){\circle*{0.700000}}
\put(23.000000,31.667093){\circle*{0.700000}}
\put(24.000000,31.405720){\circle*{0.700000}}
\put(25.000000,31.479853){\circle*{0.700000}}
\put(26.000000,31.305040){\circle*{0.700000}}
\put(27.000000,30.890306){\circle*{0.700000}}
\put(28.000000,30.853813){\circle*{0.700000}}
\put(29.000000,30.493000){\circle*{0.700000}}
\put(30.000000,30.284281){\circle*{0.700000}}
\put(31.000000,30.338066){\circle*{0.700000}}
\put(32.000000,30.207506){\circle*{0.700000}}
\put(33.000000,29.816093){\circle*{0.700000}}
\put(34.000000,29.336720){\circle*{0.700000}}
\put(35.000000,29.218639){\circle*{0.700000}}
\put(36.000000,29.367813){\circle*{0.700000}}
\put(37.000000,29.087360){\circle*{0.700000}}
\put(38.000000,29.470480){\circle*{0.700000}}
\put(39.000000,29.490000){\circle*{0.700000}}
\put(40.000000,29.482067){\circle*{0.700000}}
\put(41.000000,29.333281){\circle*{0.700000}}
\put(42.000000,29.538588){\circle*{0.700000}}
\put(43.000000,29.633654){\circle*{0.700000}}
\put(44.000000,30.085213){\circle*{0.700000}}
\put(45.000000,30.170254){\circle*{0.700000}}
\put(46.000000,30.119440){\circle*{0.700000}}
\put(47.000000,30.119667){\circle*{0.700000}}
\put(48.000000,30.053413){\circle*{0.700000}}
\put(49.000000,30.242519){\circle*{0.700000}}
\put(50.000000,30.103399){\circle*{0.700000}}
\put(51.000000,29.977880){\circle*{0.700000}}
\put(52.000000,29.833107){\circle*{0.700000}}
\put(53.000000,29.727959){\circle*{0.700000}}
\put(54.000000,29.592680){\circle*{0.700000}}
\put(55.000000,29.968201){\circle*{0.700000}}
\put(56.000000,29.930492){\circle*{0.700000}}
\put(57.000000,30.357693){\circle*{0.700000}}
\put(58.000000,30.132853){\circle*{0.700000}}
\put(59.000000,29.834654){\circle*{0.700000}}
\put(60.000000,29.573895){\circle*{0.700000}}
\put(0.000000,40.000000){\circle{0.700000}}
\put(1.000000,37.409561){\circle{0.700000}}
\put(2.000000,36.060867){\circle{0.700000}}
\put(3.000000,35.220333){\circle{0.700000}}
\put(4.000000,34.342052){\circle{0.700000}}
\put(5.000000,33.964119){\circle{0.700000}}
\put(6.000000,33.472813){\circle{0.700000}}
\put(7.000000,32.730412){\circle{0.700000}}
\put(8.000000,32.290028){\circle{0.700000}}
\put(9.000000,31.561613){\circle{0.700000}}
\put(10.000000,30.597000){\circle{0.700000}}
\put(11.000000,30.186066){\circle{0.700000}}
\put(12.000000,29.861799){\circle{0.700000}}
\put(13.000000,29.625214){\circle{0.700000}}
\put(14.000000,29.248493){\circle{0.700000}}
\put(15.000000,28.300320){\circle{0.700000}}
\put(16.000000,27.643961){\circle{0.700000}}
\put(17.000000,25.734667){\circle{0.700000}}
\put(18.000000,24.648134){\circle{0.700000}}
\put(19.000000,25.245333){\circle{0.700000}}
\put(20.000000,26.661068){\circle{0.700000}}
\put(21.000000,26.052401){\circle{0.700000}}
\put(22.000000,25.185600){\circle{0.700000}}
\put(23.000000,24.932667){\circle{0.700000}}
\put(24.000000,24.419065){\circle{0.700000}}
\put(25.000000,24.380268){\circle{0.700000}}
\put(26.000000,20.732399){\circle{0.700000}}
\put(27.000000,20.491199){\circle{0.700000}}
\put(28.000000,21.534401){\circle{0.700000}}
\put(29.000000,22.638933){\circle{0.700000}}
\put(30.000000,23.108934){\circle{0.700000}}
\put(31.000000,18.570000){\circle{0.700000}}
\put(32.000000,15.562399){\circle{0.700000}}
\put(33.000000,18.514935){\circle{0.700000}}
\put(34.000000,21.829601){\circle{0.700000}}
\put(35.000000,22.093733){\circle{0.700000}}
\put(36.000000,21.153999){\circle{0.700000}}
\put(37.000000,21.584400){\circle{0.700000}}
\put(38.000000,19.708000){\circle{0.700000}}
\put(39.000000,17.799868){\circle{0.700000}}
\put(40.000000,20.706800){\circle{0.700000}}
\put(41.000000,20.701866){\circle{0.700000}}
\put(42.000000,19.636000){\circle{0.700000}}
\put(43.000000,22.130133){\circle{0.700000}}
\put(44.000000,16.011467){\circle{0.700000}}
\put(45.000000,7.891868){\circle{0.700000}}
\put(46.000000,16.024000){\circle{0.700000}}
\put(0.000000,40.000000){\line(1,0){0.700000}}
\put(0.000000,40.000000){\line(-1,0){0.700000}}
\put(0.000000,40.000000){\line(0,1){0.700000}}
\put(0.000000,40.000000){\line(0,-1){0.700000}}
\put(1.000000,35.711601){\line(1,0){0.700000}}
\put(1.000000,35.711601){\line(-1,0){0.700000}}
\put(1.000000,35.711601){\line(0,1){0.700000}}
\put(1.000000,35.711601){\line(0,-1){0.700000}}
\put(2.000000,33.570572){\line(1,0){0.700000}}
\put(2.000000,33.570572){\line(-1,0){0.700000}}
\put(2.000000,33.570572){\line(0,1){0.700000}}
\put(2.000000,33.570572){\line(0,-1){0.700000}}
\put(3.000000,31.802000){\line(1,0){0.700000}}
\put(3.000000,31.802000){\line(-1,0){0.700000}}
\put(3.000000,31.802000){\line(0,1){0.700000}}
\put(3.000000,31.802000){\line(0,-1){0.700000}}
\put(4.000000,30.091013){\line(1,0){0.700000}}
\put(4.000000,30.091013){\line(-1,0){0.700000}}
\put(4.000000,30.091013){\line(0,1){0.700000}}
\put(4.000000,30.091013){\line(0,-1){0.700000}}
\put(5.000000,28.435040){\line(1,0){0.700000}}
\put(5.000000,28.435040){\line(-1,0){0.700000}}
\put(5.000000,28.435040){\line(0,1){0.700000}}
\put(5.000000,28.435040){\line(0,-1){0.700000}}
\put(6.000000,26.927586){\line(1,0){0.700000}}
\put(6.000000,26.927586){\line(-1,0){0.700000}}
\put(6.000000,26.927586){\line(0,1){0.700000}}
\put(6.000000,26.927586){\line(0,-1){0.700000}}
\put(7.000000,26.195333){\line(1,0){0.700000}}
\put(7.000000,26.195333){\line(-1,0){0.700000}}
\put(7.000000,26.195333){\line(0,1){0.700000}}
\put(7.000000,26.195333){\line(0,-1){0.700000}}
\put(8.000000,25.391201){\line(1,0){0.700000}}
\put(8.000000,25.391201){\line(-1,0){0.700000}}
\put(8.000000,25.391201){\line(0,1){0.700000}}
\put(8.000000,25.391201){\line(0,-1){0.700000}}
\put(9.000000,23.699600){\line(1,0){0.700000}}
\put(9.000000,23.699600){\line(-1,0){0.700000}}
\put(9.000000,23.699600){\line(0,1){0.700000}}
\put(9.000000,23.699600){\line(0,-1){0.700000}}
\put(10.000000,23.788000){\line(1,0){0.700000}}
\put(10.000000,23.788000){\line(-1,0){0.700000}}
\put(10.000000,23.788000){\line(0,1){0.700000}}
\put(10.000000,23.788000){\line(0,-1){0.700000}}
\put(11.000000,24.290266){\line(1,0){0.700000}}
\put(11.000000,24.290266){\line(-1,0){0.700000}}
\put(11.000000,24.290266){\line(0,1){0.700000}}
\put(11.000000,24.290266){\line(0,-1){0.700000}}
\put(12.000000,21.289734){\line(1,0){0.700000}}
\put(12.000000,21.289734){\line(-1,0){0.700000}}
\put(12.000000,21.289734){\line(0,1){0.700000}}
\put(12.000000,21.289734){\line(0,-1){0.700000}}
\put(13.000000,22.005867){\line(1,0){0.700000}}
\put(13.000000,22.005867){\line(-1,0){0.700000}}
\put(13.000000,22.005867){\line(0,1){0.700000}}
\put(13.000000,22.005867){\line(0,-1){0.700000}}
\put(14.000000,19.124666){\line(1,0){0.700000}}
\put(14.000000,19.124666){\line(-1,0){0.700000}}
\put(14.000000,19.124666){\line(0,1){0.700000}}
\put(14.000000,19.124666){\line(0,-1){0.700000}}
\put(15.000000,10.238266){\line(1,0){0.700000}}
\put(15.000000,10.238266){\line(-1,0){0.700000}}
\put(15.000000,10.238266){\line(0,1){0.700000}}
\put(15.000000,10.238266){\line(0,-1){0.700000}}
\put(16.000000,15.290000){\line(1,0){0.700000}}
\put(16.000000,15.290000){\line(-1,0){0.700000}}
\put(16.000000,15.290000){\line(0,1){0.700000}}
\put(16.000000,15.290000){\line(0,-1){0.700000}}
\put(17.000000,14.962800){\line(1,0){0.700000}}
\put(17.000000,14.962800){\line(-1,0){0.700000}}
\put(17.000000,14.962800){\line(0,1){0.700000}}
\put(17.000000,14.962800){\line(0,-1){0.700000}}
\put(18.000000,14.584801){\line(1,0){0.700000}}
\put(18.000000,14.584801){\line(-1,0){0.700000}}
\put(18.000000,14.584801){\line(0,1){0.700000}}
\put(18.000000,14.584801){\line(0,-1){0.700000}}
\put(19.000000,15.887334){\line(1,0){0.700000}}
\put(19.000000,15.887334){\line(-1,0){0.700000}}
\put(19.000000,15.887334){\line(0,1){0.700000}}
\put(19.000000,15.887334){\line(0,-1){0.700000}}
\put(20.000000,13.277333){\line(1,0){0.700000}}
\put(20.000000,13.277333){\line(-1,0){0.700000}}
\put(20.000000,13.277333){\line(0,1){0.700000}}
\put(20.000000,13.277333){\line(0,-1){0.700000}}
\put(21.000000,13.221467){\line(1,0){0.700000}}
\put(21.000000,13.221467){\line(-1,0){0.700000}}
\put(21.000000,13.221467){\line(0,1){0.700000}}
\put(21.000000,13.221467){\line(0,-1){0.700000}}
\put(22.000000,12.626400){\line(1,0){0.700000}}
\put(22.000000,12.626400){\line(-1,0){0.700000}}
\put(22.000000,12.626400){\line(0,1){0.700000}}
\put(22.000000,12.626400){\line(0,-1){0.700000}}
{\linethickness{0.25mm}
\put(  0,  0){\line(1,0){\xw}}
\put(  0,\yw){\line(1,0){\xw}}
\put(\xw,  0){\line(0,1){\yw}}
\put(  0,  0){\line(0,1){\yw}} }
\put(0.000000,0){\line(0,1){1}}
\put(20.000000,0){\line(0,1){1}}
\put(40.000000,0){\line(0,1){1}}
\put(60.000000,0){\line(0,1){1}}
\put(0.000000,-3.5){0}
\put(17.900000,-3.5){200}
\put(37.900000,-3.5){400}
\put(57.900000,-3.5){600}
\put(0,40.000){\line(1,0){2}}
\put(\xw,40.000){\line(-1,0){2}}
\put(0,39.390){\line(1,0){1}}
\put(\xw,39.390){\line(-1,0){1}}
\put(0,38.708){\line(1,0){1}}
\put(\xw,38.708){\line(-1,0){1}}
\put(0,37.935){\line(1,0){1}}
\put(\xw,37.935){\line(-1,0){1}}
\put(0,37.042){\line(1,0){1}}
\put(\xw,37.042){\line(-1,0){1}}
\put(0,35.986){\line(1,0){1}}
\put(\xw,35.986){\line(-1,0){1}}
\put(0,34.694){\line(1,0){1}}
\put(\xw,34.694){\line(-1,0){1}}
\put(0,33.028){\line(1,0){1}}
\put(\xw,33.028){\line(-1,0){1}}
\put(0,30.680){\line(1,0){1}}
\put(\xw,30.680){\line(-1,0){1}}
\put(0,26.667){\line(1,0){2}}
\put(\xw,26.667){\line(-1,0){2}}
\put(0,26.057){\line(1,0){1}}
\put(\xw,26.057){\line(-1,0){1}}
\put(0,25.375){\line(1,0){1}}
\put(\xw,25.375){\line(-1,0){1}}
\put(0,24.601){\line(1,0){1}}
\put(\xw,24.601){\line(-1,0){1}}
\put(0,23.709){\line(1,0){1}}
\put(\xw,23.709){\line(-1,0){1}}
\put(0,22.653){\line(1,0){1}}
\put(\xw,22.653){\line(-1,0){1}}
\put(0,21.361){\line(1,0){1}}
\put(\xw,21.361){\line(-1,0){1}}
\put(0,19.695){\line(1,0){1}}
\put(\xw,19.695){\line(-1,0){1}}
\put(0,17.347){\line(1,0){1}}
\put(\xw,17.347){\line(-1,0){1}}
\put(0,13.333){\line(1,0){2}}
\put(\xw,13.333){\line(-1,0){2}}
\put(0,12.723){\line(1,0){1}}
\put(\xw,12.723){\line(-1,0){1}}
\put(0,12.041){\line(1,0){1}}
\put(\xw,12.041){\line(-1,0){1}}
\put(0,11.268){\line(1,0){1}}
\put(\xw,11.268){\line(-1,0){1}}
\put(0,10.375){\line(1,0){1}}
\put(\xw,10.375){\line(-1,0){1}}
\put(0, 9.320){\line(1,0){1}}
\put(\xw, 9.320){\line(-1,0){1}}
\put(0, 8.027){\line(1,0){1}}
\put(\xw, 8.027){\line(-1,0){1}}
\put(0, 6.362){\line(1,0){1}}
\put(\xw, 6.362){\line(-1,0){1}}
\put(0, 4.014){\line(1,0){1}}
\put(\xw, 4.014){\line(-1,0){1}}
\put(-5,38.5){1}
\put(-8,25.1){$10^{-1}$}
\put(-8,11.8){$10^{-2}$}
\put(-8,-1.5){$10^{-3}$}
\end{picture}
\end{figure}

Finally we consider the $\beta$ dependence of
$\rho(t)$ and $\tau_{int}$.

We show the behaviors of $\rho(t)$ for different values of
$\beta$ in Fig.4(a).
We can see long range autocorrelations at $\beta$=6.33 and 6.43,
whereas rather short range ones at $\beta$=6.10 and 6.80.

The  autocorrelation time as a function of $\beta$ is displayed in Fig.4(b).
$\tau_{int}$ seems to diverge around some $\beta$,
associated with the deconfinement transition.
This indicates that around this $\beta_{\tau}$
we must be careful to evaluate the
statistical average of physical observables
as we stressed in the beginning of this paper.

 From these results, it is clear that if we want to stay in the confinement
regime,
we must stay below $\beta_{\tau}$ = 6.34 $\pm$ 0.03 on a $16^4$ lattice.

Combining with renormalization group analysis at $\beta$=6.8 on
$32^4$ lattice indicating $\Delta\beta=0.53 \pm .02$
(see Ref.5),
we must stay at $\beta \leq 6.87 \pm 0.05$ on a $32^4$ lattice.

\section*{ACKNOWLEDGEMENTS}

We are indebted to M.Ikesaka, Y.Inada, K.Inoue,
M.Ishii, T.Saito, T.Shimizu and H.Shiraishi
at the Fujitsu parallel computing research facilities
for their valuable comments on parallel computing.

\begin{figure}[t]
\unitlength 1.0mm
\begin{picture}(60,60)(-10,-10)
\put(30,-8){$\beta$}
\put(-8,38){\large $\tau_{int}$}
\put(-10,-16){Fig.4(b)\ \ {\large $\tau_{int}$} as a function of $\beta$ at
$K$=9.}
\def\xw{60.000000} \def\yw{40.000000}
\put(4.285710,2.691553){\circle*{1.000000}}
\put(8.571420,3.481800){\circle*{1.000000}}
\put(10.714285,3.316373){\circle*{1.000000}}
\put(14.999995,11.989400){\circle*{1.000000}}
\put(16.285719,25.698267){\circle*{1.000000}}
\put(17.142860,22.217600){\circle*{1.000000}}
\put(18.428562,29.287066){\circle*{1.000000}}
\put(19.285706,15.778466){\circle*{1.000000}}
\put(20.571428,27.478867){\circle*{1.000000}}
\put(21.642860,7.923000){\circle*{1.000000}}
\put(21.428570,16.034733){\circle*{1.000000}}
\put(21.857130,23.440001){\circle*{1.000000}}
\put(22.285711,18.478800){\circle*{1.000000}}
\put(22.714273,9.348933){\circle*{1.000000}}
\put(23.571415,5.581927){\circle*{1.000000}}
\put(24.428556,4.504500){\circle*{1.000000}}
\put(29.999990,2.649673){\circle*{1.000000}}
\put(38.571430,2.984753){\circle*{1.000000}}
\put(47.142849,1.793520){\circle*{1.000000}}
\put(55.714272,4.077613){\circle*{1.000000}}
\put(4.285710,2.691553){\line(0,1){0.891033}}
\put(4.285710,2.691553){\line(0,-1){0.891033}}
\put(8.571420,3.481800){\line(0,1){1.045253}}
\put(8.571420,3.481800){\line(0,-1){1.045253}}
\put(10.714285,3.316373){\line(0,1){0.687020}}
\put(10.714285,3.316373){\line(0,-1){0.687020}}
\put(14.999995,11.989400){\line(0,1){8.593334}}
\put(14.999995,11.989400){\line(0,-1){8.593334}}
\put(16.285719,25.698267){\line(0,1){15.576134}}
\put(16.285719,25.698267){\line(0,-1){15.576134}}
\put(17.142860,22.217600){\line(0,1){10.066200}}
\put(17.142860,22.217600){\line(0,-1){10.066200}}
\put(18.428562,29.287066){\line(0,1){14.168200}}
\put(18.428562,29.287066){\line(0,-1){14.168200}}
\put(19.285706,15.778466){\line(0,1){7.218200}}
\put(19.285706,15.778466){\line(0,-1){7.218200}}
\put(20.571428,27.478867){\line(0,1){9.124067}}
\put(20.571428,27.478867){\line(0,-1){9.124067}}
\put(21.642860,7.923000){\line(0,1){3.041660}}
\put(21.642860,7.923000){\line(0,-1){3.041660}}
\put(21.428570,16.034733){\line(0,1){8.507733}}
\put(21.428570,16.034733){\line(0,-1){8.507733}}
\put(21.857130,23.440001){\line(0,1){8.090067}}
\put(21.857130,23.440001){\line(0,-1){8.090067}}
\put(22.285711,18.478800){\line(0,1){13.389867}}
\put(22.285711,18.478800){\line(0,-1){13.389867}}
\put(22.714273,9.348933){\line(0,1){5.648647}}
\put(22.714273,9.348933){\line(0,-1){5.648647}}
\put(23.571415,5.581927){\line(0,1){1.772727}}
\put(23.571415,5.581927){\line(0,-1){1.772727}}
\put(24.428556,4.504500){\line(0,1){1.661587}}
\put(24.428556,4.504500){\line(0,-1){1.661587}}
\put(29.999990,2.649673){\line(0,1){0.441964}}
\put(29.999990,2.649673){\line(0,-1){0.441964}}
\put(38.571430,2.984753){\line(0,1){0.698033}}
\put(38.571430,2.984753){\line(0,-1){0.698033}}
\put(47.142849,1.793520){\line(0,1){0.289063}}
\put(47.142849,1.793520){\line(0,-1){0.289063}}
\put(55.714272,4.077613){\line(0,1){2.558780}}
\put(55.714272,4.077613){\line(0,-1){2.558780}}
{\linethickness{0.25mm}
\put(  0,  0){\line(1,0){\xw}}
\put(  0,\yw){\line(1,0){\xw}}
\put(\xw,  0){\line(0,1){\yw}}
\put(  0,  0){\line(0,1){\yw}} }
\put(4.285710,0){\line(0,1){1}}
\put(12.857130,0){\line(0,1){1}}
\put(21.428570,0){\line(0,1){1}}
\put(29.999990,0){\line(0,1){1}}
\put(38.571430,0){\line(0,1){1}}
\put(47.142849,0){\line(0,1){1}}
\put(55.714272,0){\line(0,1){1}}
\put(2.085710,-3.5){6.0}
\put(10.657130,-3.5){6.2}
\put(19.228570,-3.5){6.4}
\put(27.799990,-3.5){6.6}
\put(36.371430,-3.5){6.8}
\put(44.942849,-3.5){7.0}
\put(53.514272,-3.5){7.2}
\put(0,0.000000){\line(1,0){1}}
\put(\xw,0.000000){\line(-1,0){1}}
\put(0,6.666667){\line(1,0){1}}
\put(\xw,6.666667){\line(-1,0){1}}
\put(0,13.333333){\line(1,0){1}}
\put(\xw,13.333333){\line(-1,0){1}}
\put(0,20.000000){\line(1,0){1}}
\put(\xw,20.000000){\line(-1,0){1}}
\put(0,26.666666){\line(1,0){1}}
\put(\xw,26.666666){\line(-1,0){1}}
\put(0,33.333332){\line(1,0){1}}
\put(\xw,33.333332){\line(-1,0){1}}
\put(-1.900000,-1.500000){0}
\put(-5.700000,5.166667){100}
\put(-5.700000,11.833333){200}
\put(-5.700000,18.500000){300}
\put(-5.700000,25.166666){400}
\put(-5.700000,31.833332){500}
\end{picture}
\end{figure}

\end{document}